\documentclass[aps,prb,preprint,groupedaddress,showpacs]{revtex4}
\usepackage{graphicx}
\usepackage{amssymb}
\usepackage{amsmath}
\usepackage{latexsym}
\usepackage{ulem}
\usepackage{color}
\usepackage{dcolumn}
\usepackage{subfigure}

\begin{document}

\title{Spontaneous magnetization in presence of 
nanoscale inhomogeneities in diluted magnetic systems}

\author{Akash Chakraborty}
\affiliation{
Institut f\"ur Theoretische Festk\"orperphysik, Karlsruhe Institute of Technology, 76128 Karlsruhe, Germany
}
\affiliation{
School of Engineering and Science, Jacobs University Bremen, Campus Ring 1, 28759 Bremen, Germany
}
\author{Paul Wenk}
\affiliation{
Institut I - Theoretische Physik, Universit\"at Regensburg, 93040 Regensburg, Germany
}

\author{Richard Bouzerar}
\affiliation{
Institut N\'eel, CNRS, D\'epartement MCBT, 25 avenue des Martyrs, B.P. 166, 38042 Grenoble Cedex 09, France
}

\author{Georges Bouzerar}
\affiliation{
Institut N\'eel, CNRS, D\'epartement MCBT, 25 avenue des Martyrs, B.P. 166, 38042 Grenoble Cedex 09, France
}
\affiliation{
School of Engineering and Science, Jacobs University Bremen, Campus Ring 1, 28759 Bremen, Germany
}

\date{\today}

\begin{abstract}
The presence of nanoscale inhomogeneities has been experimentally evidenced in several diluted magnetic systems, which in turn often leads to interesting physical phenomena. 
However, a proper theoretical understanding of the underlying physics is lacking in most of the cases. Here we present a detailed and comprehensive theoretical study of the 
effects of nanoscale inhomogeneities on the temperature dependent spontaneous magnetization in diluted magnetic systems, which is found to exhibit an unusual and 
unconventional behavior. The effects of impurity clustering on the magnetization 
response have hardly been studied until now. We show that nano-sized clusters of magnetic impurities can lead to drastic effects on the magnetization compared to that of 
homogeneously diluted compounds. The anomalous nature of the magnetization curves strongly depends on the relative concentration of 
the inhomogeneities as well as the effective range of the exchange interactions. In addition we also provide a systematic discussion of the nature of the distributions of the 
local magnetizations. 
\end{abstract}


\maketitle

\ Interest in diluted magnetic systems, such as diluted magnetic semiconductors (DMSs)\cite{timm,jungwirth,satormp} and diluted magnetic oxides (DMOs)\cite{fukumura,chambers}, 
has continued to surge over the past couple of 
decades owing to their huge potential for spintronics applications. The prime requisite of room-temperature ferromagnetism for feasible spintronics devices has led to  
Curie temperature ($T_C$) being the primary focus of interest in these diluted magnetic materials. Contrary to a longstanding belief of defect and inhomogeneity free 
samples leading to high Curie temperatures, recent studies suggest otherwise. Experimental studies have revealed the formation and existence of magnetic clusters on the nanoscale 
order, in materials like (Ge,Mn)\cite{jamet} and (Zn,Co)O\cite{jedrecy}, which in turn gave rise to very high $T_C$'s ($\ge$300 K). In fact, very recently we have theoretically
shown that incorporating magnetic nanoclusters, can lead to drastically high critical temperatures (often above room-temperature) in diluted magnetic systems with effective 
short-ranged exchange interactions\cite{akash}. Now, another important aspect which can help to 
further reveal the physical intricacies of these strongly disordered and complex systems is the temperature dependence of the spontaneous magnetization $M(T)$. Among the many 
interesting features that magnetization possesses, some worth mentioning are the convex or concave nature of the $M(T)$ curves and their critical behavior close to the transition 
point.

\ In the particular case of DMSs, one of the very first observations of ferromagnetism in (In,Mn)As\cite{ohno}, revealed a surprising non-mean-field like behavior of the spontaneous 
magnetization with temperature. The experimentally determined magnetization curve had an unusual outward concave-like shape which is in stark contrast to the typical convex 
behavior obtained within the standard Weiss mean-field theory\cite{ashcroft}, as well as that observed in conventional ferromagnetic materials. The (In,Mn)As samples 
studied in Ref.\onlinecite{ohno} were reported to be insulating. Similar concave $M(T)$ behavior was also observed in insulating samples of Ge$_{1-x}$Mn$_{x}$ determined 
by superconducting quantum interference device (SQUID) magnetometry and magnetotransport measurements\cite{ydpark}. Theoretical predictions, based on a percolation 
transition of bound magnetic polarons in the strongly localized regime\cite{kaminski}, were made to explain this non-mean-field like magnetization behavior. Similar 
magnetization behavior in DMS systems, in the insulating regime, was also predicted by other theoretical studies based on numerical calculations\cite{mayr,kennett,berciu1}. 
The deviation in the $M(T)$ behavior from the standard Brillouin-function shape was believed to be partly due to the small carrier density compared to the localized spin 
density, as well as due to the wide distribution of the exchange interactions and hopping integrals\cite{berciu1}. On the other hand, in metallic DMSs, for example 
Ga$_{1-x}$Mn$_{x}$As for $x$=0.05$-$0.10, the magnetization behavior is found to decrease almost linearly 
with temperature\cite{beschoten,mathieu}. This behavior is somewhat 
intermediate between the concave-like $M(T)$ curves in the insulating regime and the classic convex magnetization. However, experimental studies suggest that annealing 
treatments can have a strong effect on the nature of the magnetization behavior\cite{edmonds,potashnik}. During annealing the magnetization curves are found to 
become more convex and Brillouin-function-like compared to their linear behavior before annealing. This anomalous behavior of the magnetization highlights the 
importance of disorder in these systems. 

\ However, the effects of correlations in disorder or impurity clustering have hardly been considered barring a few cases. In 
Ref.\onlinecite{timm1}, the authors reported that correlated defects lead to ``mean-field-like'' magnetization curves in (Ga,Mn)As. In another theoretical 
study\cite{sanyal} on Co doped ZnO, inhomogeneous phases were shown to be responsible for high Curie temperatures in comparison to the homogeneous samples.
Now a majority of the existing theoretical studies are based on mean-field-like approaches, which is known to be inadequate to treat thermal and/or transverse fluctuations 
and disorder reliably in these systems. In Ref.\onlinecite {sarma}, the authors 
have used some complementary theoretical approaches in addition to the mean-field theory, which include the dynamical mean-field theory (DMFT), to study the temperature dependent 
magnetization in doped magnetic semiconductors. Nonetheless, the effects of clustered defects were not taken into account. The presence of nanoscale inhomogeneities can give rise 
to very interesting and new physics in these diluted systems, as was seen in the case of the Curie temperatures\cite{akash}. 

\ The primary objective of the current manuscript is 
to investigate the effects of these inhomogeneities on the spontaneous magnetization behavior. Here we first study 
the nature of the magnetization in homogeneously diluted systems with no correlation in impurity positions, and then extend this to systems containing clusters of magnetic 
impurities. We observe very interesting as well as strong deviations from the homogeneous magnetization curves. A non-trivial nature of the spontaneous magnetization in these 
inhomogeneous diluted systems is found to strongly depend on the relative concentration of the inhomogeneities as well as the effective range of the magnetic exchange interactions.

\ For the sake of simplicity we have assumed here a simple cubic lattice with periodic boundary conditions. We fix the total concentration of impurities in the system to 
$x$=0.07 and the total number of impurities in the system is denoted by $N_{imp}$. The inhomogeneities 
are assumed to be of spherical shape of radii $r_0$. The concentration of impurities inside each nanosphere is denoted by $x_{in}$. To avoid additional parameters, we restrict 
ourselves to nanospheres of fixed radii $r_0$=2$a$ ($a$ is the lattice spacing) and $x_{in}$=0.8. The concentration of nanospheres in the system is defined by 
$x_{ns}$=$N_{S}$/$N$, where $N_{S}$ is the total number of sites included in all the nanospheres and $N$=$L^3$ is the total number of sites. We define a variable 
$P_N$=($x_{in}$/$x$)$x_{ns}$, which denotes the total fraction of impurities contained within the nanospheres. For a particular configuration,  
the nanospheres are distributed in a random fashion on the lattice, the only restriction being to avoid any overlap with each other.

In order to calculate the spontaneous magnetization, we start with the effective diluted Heisenberg Hamiltonian describing $N_{imp}$ interacting spins (${\bf S}_{i}$) 
randomly distributed on a lattice of $N$ sites, given by
\begin{eqnarray}
H=-\sum_{i,j} J_{ij}p_{i}p_{j} {\bf S}_{i}\cdot{\bf S}_{j}
\label{Hamiltonian}
\end{eqnarray}
where the sum $ij$ runs over all sites and the random variable $p_i$=1 if the site is occupied by an impurity or otherwise 0. We assume magnetic couplings of the form 
$J_{ij}$=$J_{0}$exp(-$\mid$$ \textbf r $$\mid$/$\lambda$), where \textbf r=\textbf r$_i$-\textbf r$_j$. Both, the choice for such couplings and 
the values of the damping parameter $\lambda$ are discussed below.

\begin{figure}[t]\centerline
{\includegraphics[width=4.0in,angle=-0]{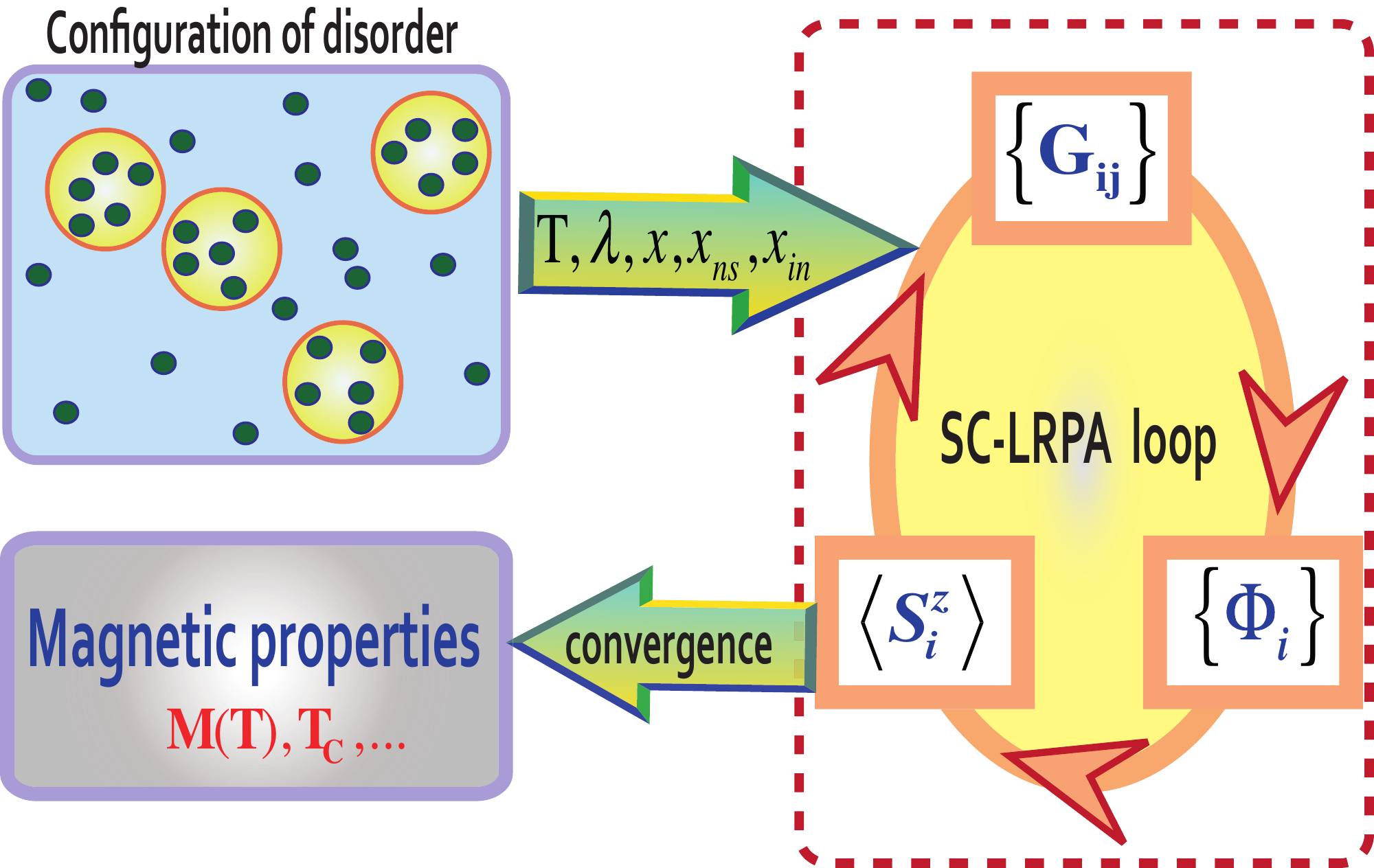}}
\caption{(Color online) Illustration of the self-consistent local random phase approximation method at a given temperature $T$.
}
\label{fig1}
\end{figure} 

\ This Hamiltonian (Eq.1) is treated within the 
self-consistent local random phase approximation (SC-LRPA) theory, which is essentially a semi-analytical approach based on finite temperature Green's functions. 
More details on the SC-LRPA theory can be found in Refs.\onlinecite{satormp,georges1}. A schematic illustration of the method is given in Fig.~\ref{fig1}.
We briefly summarize the method in the following. We  start the self-consistent loop with a fully polarized, collinear ferromagnetic ground state at $T$=0 K. The impurity
spins are assumed to be classical in this case, although the theory is valid for quantum spins as well.   Within the SC-LRPA formalism, for a given 
disorder configuration, the local magnetizations at each site $\langle {S^i_z} \rangle$ ($i$=$1,2,...,N_{imp}$) are calculated self-consistently at each temperature. The local magnetization is 
evaluated using a Callen-like expression \cite{callen}, which relates the local Green's function at site $i$ to the local magnetization at this site,
\begin{eqnarray}
\langle S^i_z \rangle=\frac{(S-\Phi_i)(1+\Phi_i)^{2S+1}+(1+S+\Phi_i){\Phi_i}^{2S+1}}{(1+\Phi_i)^{2S+1}-\Phi_i^{2S+1}}
\label{Callen}
\end{eqnarray}
where the local effective magnon occupation number is given by 
\begin{eqnarray}
\Phi_i=-\frac{1}{2\pi\langle{S^i_z}\rangle}\int_{-\infty}^{+\infty}\frac{\Im G_{ii}(\omega)}{\exp(\omega/k_{B}T)-1}d\omega
\label{magnon}
\end{eqnarray}
We define the retarded Green's function as $G_{ij}(\omega)$=$\int_{-\infty}^{\infty}G_{ij}(t)e^{i\omega t}dt$, where $G_{ij}(t)$=-$i\theta(t)\langle [S_i^+(t),S_j^-(0)]\rangle$,
 which describes the transverse spin fluctuations, and $\langle ... \rangle$ denotes the expectation value.
 By the self-consistent treatment we obtain, for a given temperature and disorder configuration, the average magnetization which is defined by 
$\langle{S_{z}^{avg}}\rangle$=$\frac {1}{N_{imp}}\sum_{i=1}^{N_{imp}}\langle{S^i_z}\rangle$. The accuracy and 
reliability of the SC-LRPA to treat thermal/transverse fluctuations and disorder/dilution have been shown on several occasions\cite{satormp,richard1}. Moreover it has 
also been applied to calculate the Curie temperatures in inhomogeneous diluted systems in a recent study\cite{akash}.

 The assumption of the magnetic interactions of the form $J_{ij}$=$J_{0}$exp(-$\mid$$ \textbf r_i$-$\textbf r_j $$\mid$/$\lambda)$ is based on the fact 
that both \textit{ab initio} studies\cite{bergqvist,satoprb} as well as model calculations\cite{akash2} have shown that the exchange couplings in III-V DMSs are 
relatively short ranged and non-oscillating in nature. This non-oscillating behavior results from the existence of a preformed impurity band (IB)\cite{sandratskii}. The 
existence/absence of an IB has been a longstanding debate for several years. But in 
our view, the still often quoted RKKY interactions in these compounds, only consistent with the valence band (VB) or perturbative picture,  
can be ruled out. As shown in Ref.\onlinecite{richard1}, only the IB scenario could explain the proximity of (Ga,Mn)As to the 
metal-insulator transition as well as the observed red shift in the optical conductivity in this compound. Furthermore, recent experimental findings\cite{dobrowolska} 
definitely corroborate the fact that the Fermi level is located within an IB in (Ga,Mn)As. Now, in (Ga,Mn)As, for about 5\% Mn, a fit of the \textit{ab initio} exchange couplings provides 
a value of $\lambda$ of the order of $a/2$. It should be noted that in the case of (Ga,Mn)N the exchange interactions are even shorter ranged. 
In the following we will consider a range of $\lambda$'s, corresponding to relatively long-ranged couplings down to shorter ranged ones, and try to analyze their effects on 
the magnetization behavior. In fact the magnetic couplings could also depend on 
the nanoscale inhomogeneities. However, computing these would require extensive calculations involving finite size analyses, systematic average over disorder configurations 
and especially diagonalizing considerably large matrices. This is beyond the scope of the present manuscript. Nevertheless, we believe that the overall nature of the 
exchange couplings, especially in wide gap compounds like (Ga,Mn)N, will remain essentially unchanged. The interesting feature which we want to
focus on in the current work is the role the inhomogeneities play in determining the nature of the magnetization curves. In the following, the average magnetization is always 
plotted as a function of the reduced temperature $T/T^*$, where $T^*$ is the temperature corresponding to the case when $\langle{S_z^{avg}}\rangle$=0.001$S$. We have chosen 
$T^*$ instead of $T_C$ as it is difficult to determine accurately the critical temperatures from the magnetization curves. However, we have checked for some cases, 
that $T^*$ is relatively close to the $T_C$ directly calculated from the semi-analytical expression (Eq.(1) in Ref.\onlinecite{akash}).

\begin{figure}[t]\centerline
{\includegraphics[width=5.0in,angle=0]{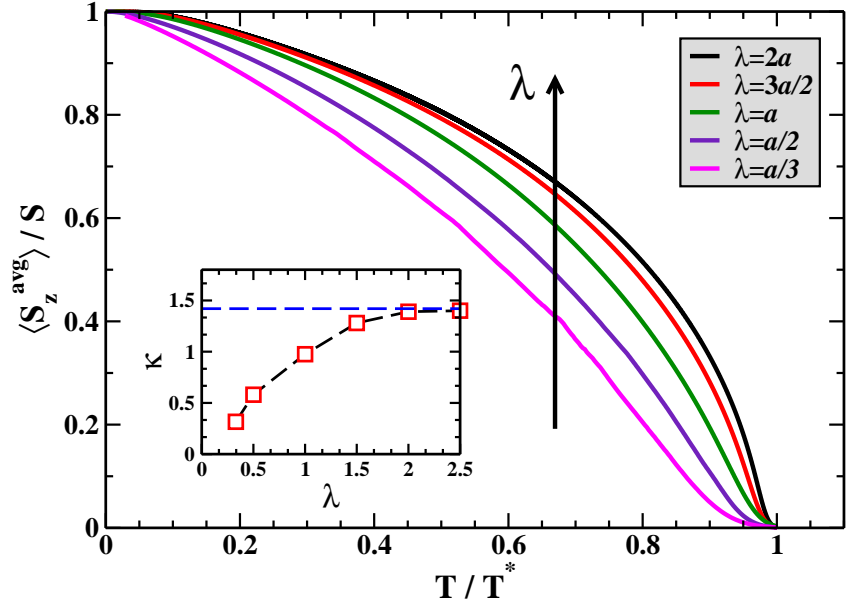}}
\caption{(Color online) Average magnetization as a function of $T$/$T^*$ for different values of $\lambda$, corresponding to the homogeneous case for a single 
configuration. Inset: Curvature of the $\langle{S_z^{avg}}\rangle$ curves as a function of $\lambda$, calculated at $T/T^*$=0.5. The blue dashed line 
 indicates the saturation value of the curvature. (Here $N$=24$^3$).
}
\label{fig2}
\end{figure} 

\ To begin with we consider the homogeneously diluted case where the magnetic impurities are randomly distributed on the lattice. Fig.~\ref{fig2} shows the average 
magnetization as a function of the reduced temperature $T/T^*$, for different values of $\lambda$. The magnetization shown corresponds to a 
single configuration of disorder and the system size is $L$=24. We observe that for relatively long-ranged couplings ($\lambda$=2$a$), the magnetization curve has a pronounced convex 
shape which is the usual behavior predicted by the mean-field theories of Weiss and Stoner\cite{ashcroft}, and observed commonly in conventional ferromagnetic materials. 
On decreasing $\lambda$, we notice that the convexity decreases and for $\lambda$=$a/3$ the magnetization is more linear-like over a broad temperature range. In fact similar 
behavior (linearity) of the magnetization was observed in metallic samples of (Ga,Mn)As by magnetic circular dichroism (MCD) studies\cite{beschoten}. This shows that the 
relatively short ranged interactions are more relevant for the case of DMS materials and also vindicates the choice of our exchange couplings. In order to have a qualitative 
idea of the relative change in the behavior of the magnetization with $\lambda$, we have plotted in  the inset of  Fig.~\ref{fig2} the curvature ($\kappa$)
 of the magnetization curves at the specific value of $T/T^*$=0.5. The curvature is defined by
$\kappa$=$\left|\frac{\partial^{2}\langle{S_{z}^{avg}}\rangle/S}{\partial u^{2}}\right| \left(1+\left[\frac{\partial\langle S_{z}^{avg}\rangle/S}{\partial u}\right]^{2}\right)^{-3/2}$, 
where $u$=$\frac{T}{T^*}$. As can be clearly seen, with increase in $\lambda$ the curvature changes significantly, increasing by almost a factor of five 
from $\lambda$=$a/3$ to $\lambda$=$2a$. For $\lambda$$\ge$2a, one sees that $\kappa$ has already saturated and the magnetization has a standard Brillouin shape.
This implies that $\lambda$$\ge$2a corresponds to the long range coupling regime. 

\begin{figure}[t]\centerline
\centering
\subfigure{
\includegraphics[width=2.2in,angle=0]{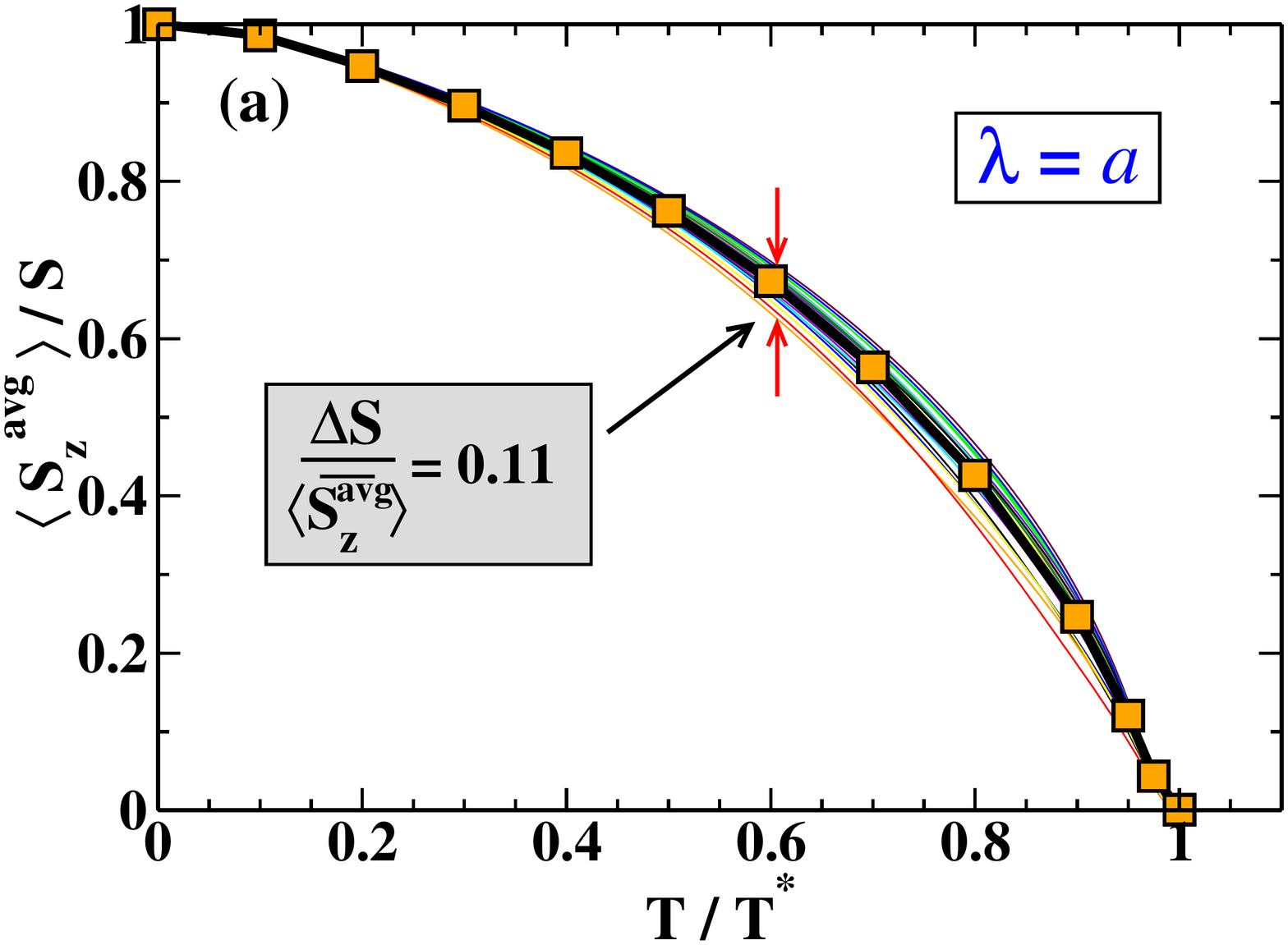}
\label{fig3a}
}
\hspace{-1.0cm}
\subfigure{
\includegraphics[width=2.2in,angle=0]{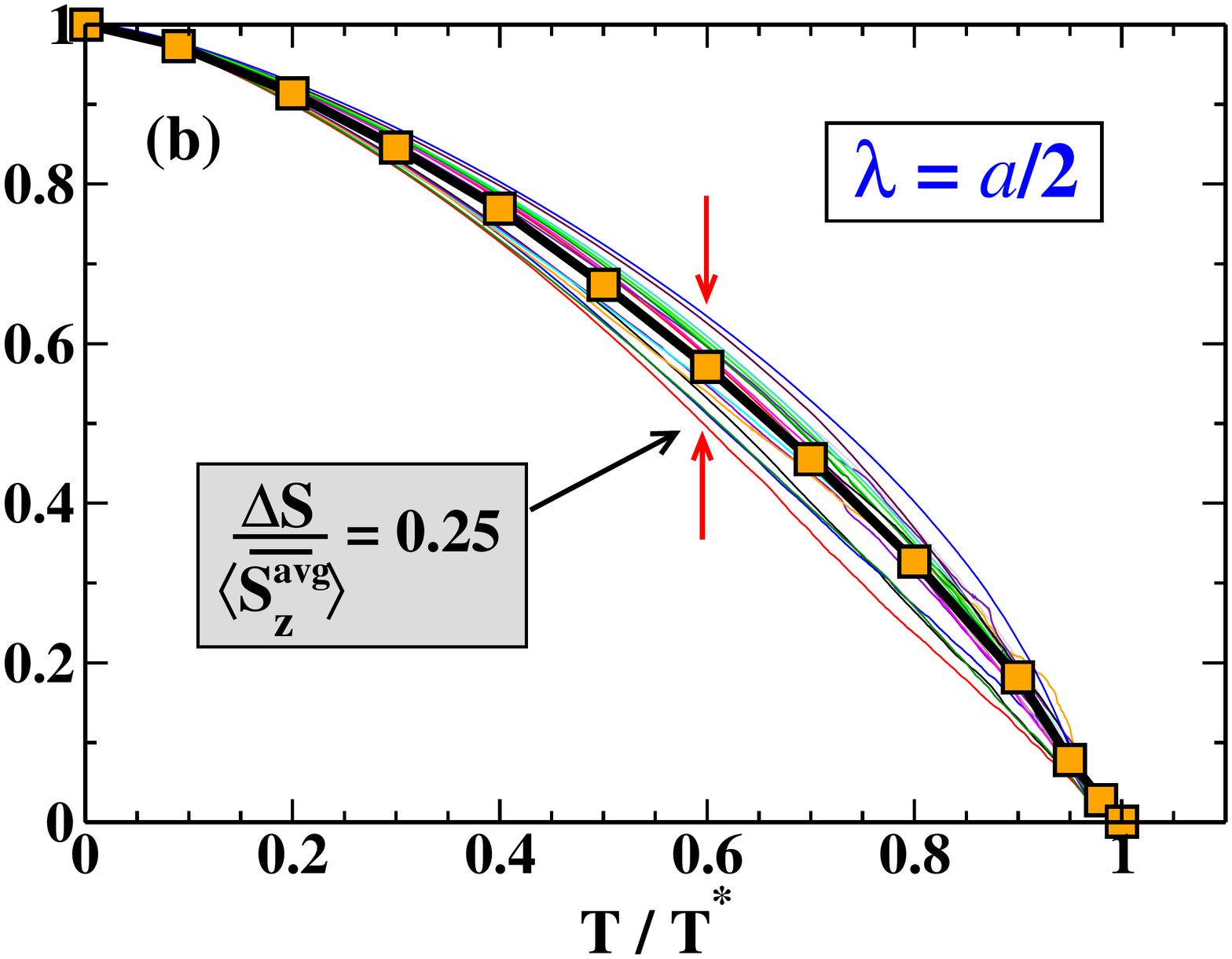}
\label{fig3b}
}
\hspace{-1.0cm}
\subfigure{
\includegraphics[width=2.2in,angle=0]{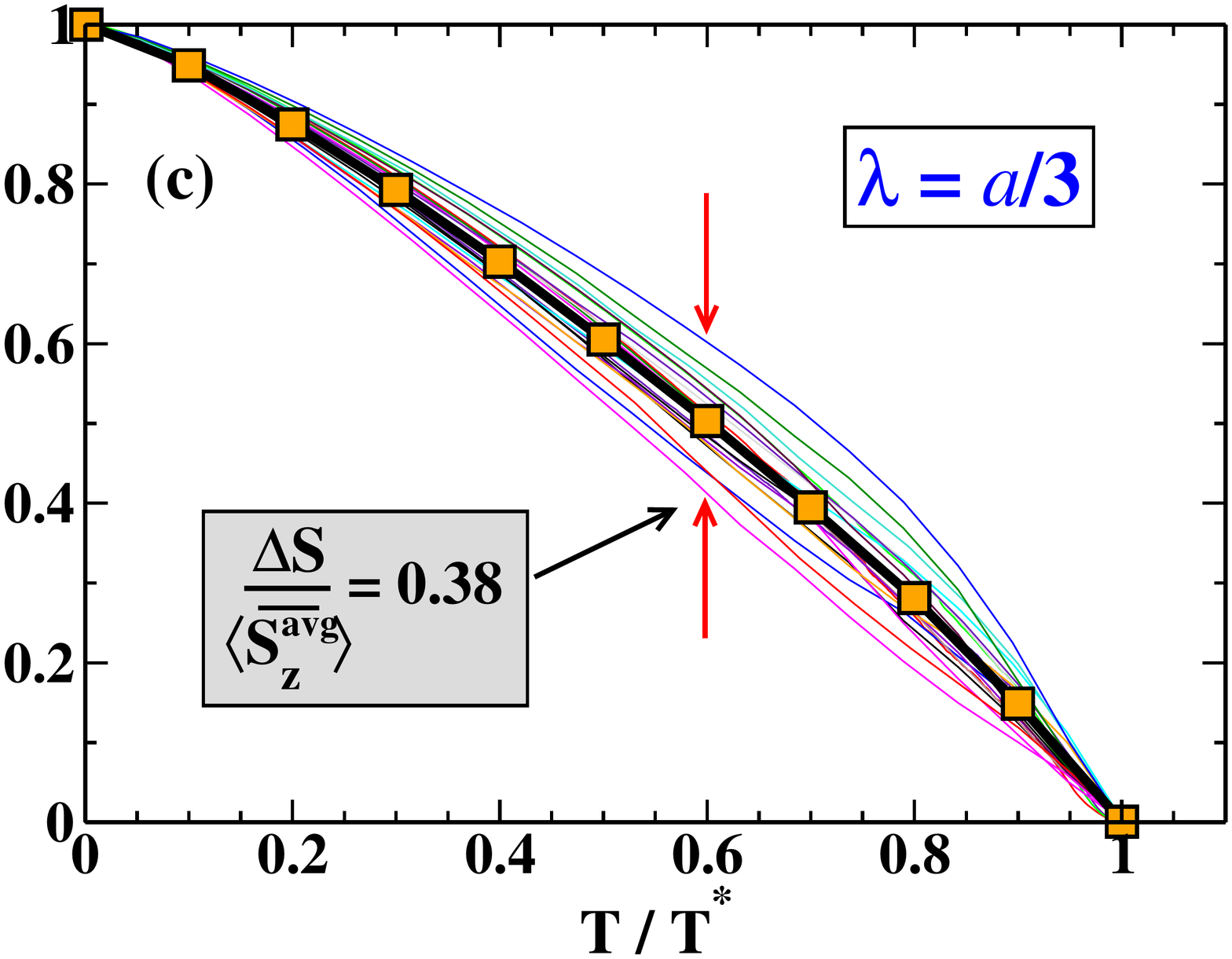}
\label{fig3c}
}
\label{fig3}
\caption[Optional caption for list of figures]{
%Caption of subfigures \ref{fig2a}, \ref{fig2b}
(Color online) Average magnetization for the homogeneous case calculated for 25 configurations, for (a)$\lambda$=$a$, (b)$\lambda$=$a/2$, and (c)$\lambda$=$a/3$. The thick 
black lines with symbols indicate the configuration averaged magnetization. $\frac{\Delta S}{\langle \overline{S_{z}^{avg}}\rangle}$ is a measure of the fluctuation at $T/T^*$=0.6. 
(Here $N$=24$^3$).
}
\end{figure}

In Fig.~\ref{fig2}, we have shown the magnetization for a single configuration of disorder. However, we know that in diluted materials the magnetic properties are often very 
sensitive to the random impurity configurations. Fig. 3 shows the average magnetization calculated for 25 configurations of disorder corresponding to three different values 
of $\lambda$. We have also calculated $\frac{\Delta S}{\langle \overline{S_{z}^{avg}}\rangle}$, which gives a measure of the fluctuations of 
the extremal magnetization curves from the configuration averaged one (${\langle \overline{S_{z}^{avg}}\rangle}$), at the particular value of $T/T^*$=0.6. For 
$\lambda$=$a$ (Fig.~\ref{fig3a}), we observe that the spontaneous magnetization is weakly  sensitive to the disorder configurations, the overall shape of the curves is unchanged. 
$\frac{\Delta S}{\langle \overline{S_{z}^{avg}}\rangle}$ is found to vary within 10\% of the configuration averaged magnetization. For $\lambda$=$2a$ (not shown here), these 
fluctuations were found to be even smaller, varying within less than 5\%. For the short ranged couplings, $\lambda$=$a/2$, 
the magnetization curves are still convex but the fluctuations are stronger now. $\frac{\Delta S}{\langle \overline{S_{z}^{avg}}\rangle}$ is more than doubled as compared to the 
intermediate range of $\lambda$=$a$.  As can be seen from Fig.~\ref{fig3b}, some of the  curves have a regular convex behavior while some are more linear in nature.  Now on further 
reducing $\lambda$ (Fig.~\ref{fig3c}), the deviations become even stronger, and a significant number of configurations exhibit a clear linear temperature dependence.
The fluctuation at $T/T^*$=0.6 increases by almost a factor four, compared to the case of $\lambda$=$a$. Even, in some cases, the magnetization profiles are slightly 
concave toward the high temperature. It should be noted that the more linear or concave magnetization curves correspond to relatively high $T_C$'s. This figure clearly 
shows that the disorder effects are significantly enhanced in the case of short-ranged interactions. The primary reason is that the probability to find regions of weakly interacting 
impurities increases significantly for the case of short-ranged interactions. This is the case in most of the DMSs, where a non-trivial behavior of the magnetization is observed.

\begin{figure}[t]\centerline
{\includegraphics[width=5.5in,angle=0]{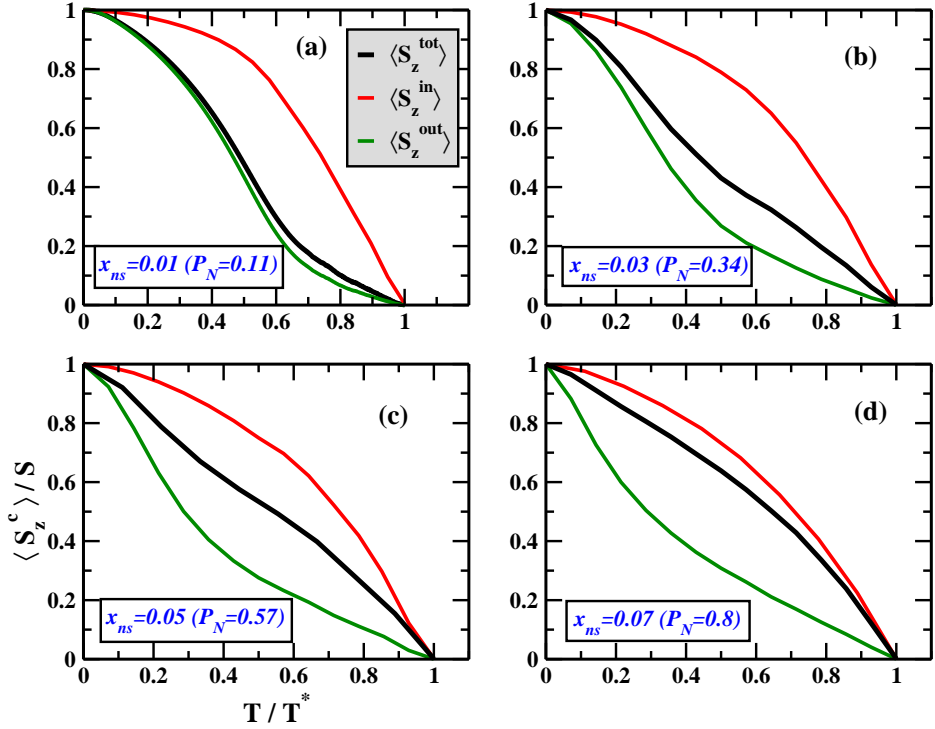}}
\caption{(Color online)$\langle{S_z^{c}}\rangle$ stands for the total average magnetization ($\langle{S_z^{tot}}\rangle$), magnetization inside the nanospheres 
($\langle{S_z^{in}}\rangle$) and outside the nanospheres ($\langle{S_z^{out}}\rangle$).  Four different concentrations of nanospheres ($x_{ns}$): (a) 0.01, (b) 0.03, (c) 0.05, 
and (d) 0.07, are considered. $P_N$ is the percentage of total impurities contained in the nanospheres.  Here $\lambda$=$a$, r$_0$=$2a$, $x_{in}$=0.8 and $N$=24$^3$. 
The x-axis is in units of $T/T^*$.
} 
\label{fig4}
\end{figure}

\ So far we have only considered homogeneously diluted systems assuming a fully random distribution of the impurities. Now we move to the case of nanoclusters of magnetic 
impurities. Fig.~\ref{fig4} shows for $\lambda$=$a$, the average magnetization of the whole system, $\langle{S_z^{tot}}\rangle$, as a function of $T/T^*$, for four 
different concentrations of nanospheres, $x_{ns}$=1\%, 3\%, 5\%, and 7\%. In addition, we have also shown the average magnetization 
inside and outside the clusters denoted by $\langle{S_z^{in}}\rangle$ and $\langle{S_z^{out}}\rangle$, respectively.  The curves shown here correspond to a single 
configuration of disorder, the variation with disorder configurations will be discussed in what follows. We immediately observe that in the presence of inhomogeneities 
the spontaneous magnetization has a non-trivial behavior and exhibits a drastically different nature when compared to the homogeneous case (Fig.~\ref{fig2}). This can be clearly seen 
even for the lowest concentration of nanospheres.  For $x_{ns}$=0.01, for which 11\% of the total impurities are inside the nanospheres, 
 $\langle{S_z^{tot}}\rangle$ decreases rapidly till about $T/T^*$$\sim$0.5, then it becomes concave and decays slowly toward the higher temperatures.
By gradually increasing the concentration of the nanospheres, an interesting change in the average magnetization behavior is observed. For $x_{ns}$=3\% ($P_N$=0.34), 
$\langle{S_z^{tot}}\rangle$ falls off less sharply at low temperature, for 5\% it is almost linear over the entire temperature range, and for 7\% it becomes more convex. Thus a 
crossover in the curvature of $\langle{S_z^{tot}}\rangle$ appears at $x_{ns}$$\approx$0.05. On the other hand, 
$\langle{S_z^{in}}\rangle$ exhibits a clear convex nature which does not change with $x_{ns}$. This indicates that the average magnetization inside the clusters remains 
almost uniform and is mainly controlled by the intra-cluster couplings. We can clearly see that the inhomogeneities have a very strong effect on the 
impurities outside the clusters.  $\langle{S_z^{out}}\rangle$ has a very pronounced concave nature
which can even be seen for relatively small $x_{ns}$. The slope at low temperatures becomes steeper with increasing concentration of nanospheres. For example, at  
$T/T^*$$\sim$0.3, $\langle{S_z^{out}}\rangle$ has a value of 0.85 for $x_{ns}$=0.01, 0.62 for $x_{ns}$=0.03, and about 0.4 for $x_{ns}$=0.05.  Similar concave behavior of the 
temperature dependent magnetization is observed in the case of some insulating DMS materials\cite{ohno}. However, in most of the cases studied until now clustering 
effects have hardly been considered. Note that we have also performed the calculations for $\lambda$=$2a$ (not shown here). In this extended coupling regime, 
it was found that the effect of inhomogeneities are very weak. $\langle{S_z^{tot}}\rangle$, $\langle{S_z^{in}}\rangle$, and $\langle{S_z^{out}}\rangle$ exhibit a convex 
nature and were found to be relatively close to each other.

\begin{figure}[t]\centerline
{\includegraphics[width=5.5in,angle=0]{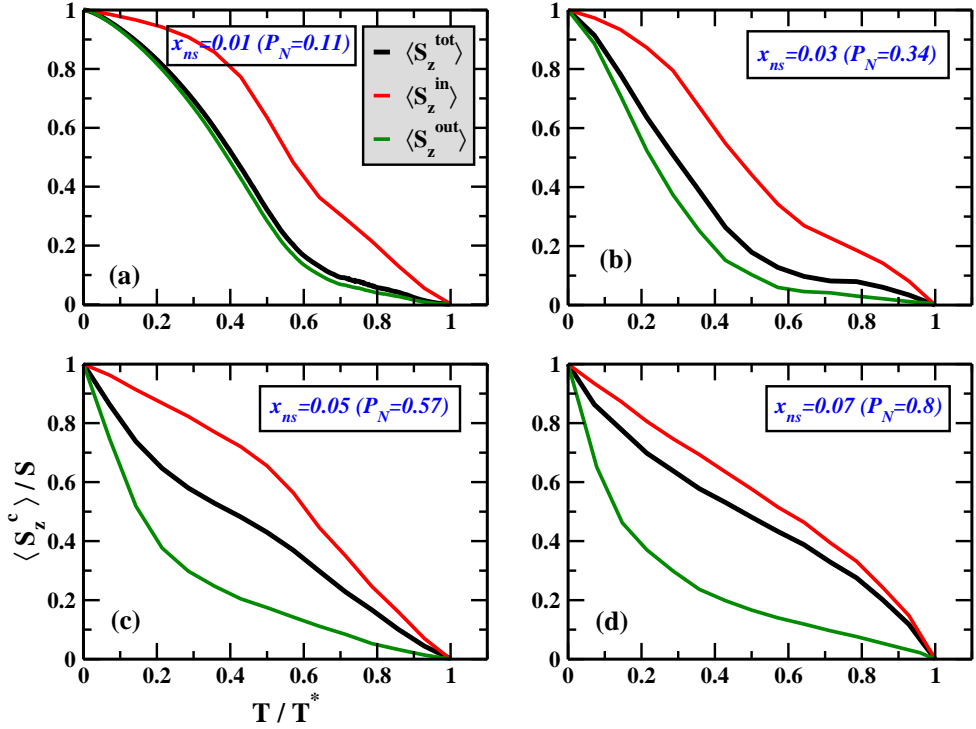}}
\caption{(Color online)$\langle{S_z^{c}}\rangle$ stands for the total average magnetization ($\langle{S_z^{tot}}\rangle$), magnetization inside the nanospheres ($\langle{S_z^{in}}\rangle$) and outside
the nanospheres ($\langle{S_z^{out}}\rangle$).  Four different concentrations of nanospheres ($x_{ns}$): (a) 0.01, (b) 0.03, (c) 0.05, and (d) 0.07, are considered. $P_N$ is the 
percentage of total impurities contained in the nanospheres. Here $\lambda$=$a/2$,
r$_0$=$2a$, $x_{in}$=0.8 and $N$=24$^3$. The x-axis is in units of $T/T^*$.
} 
\label{fig5}
\end{figure} 

\ In an earlier study, we have shown that relatively short-ranged couplings appear to have spectacular effects  on the Curie temperatures
in the presence of nanoscale inhomogeneities\cite{akash}. Short-ranged interactions are more relevant from the practical point of view. As mentioned above, the magnetic couplings in 
most III-V DMS materials ((Ga,Mn)As, (Ga,Mn)N, (Ga,Mn)P,...) are effectively short-ranged in nature. 
In Fig.~\ref{fig5} we show $\langle{S_z^{tot}}\rangle$, $\langle{S_z^{in}}\rangle$, and $\langle{S_z^{out}}\rangle$ as a function of $T/T^*$, for four 
different concentrations of nanospheres, in the case of relatively short-ranged couplings, namely $\lambda$=$a/2$. To start with, we first discuss the results for a single 
configuration of disorder. For the lowest $x_{ns}$ (Fig.~\ref{fig5}(a)), the behavior of $\langle{S_z^{tot}}\rangle$ is almost similar to that observed in the case of 
$\lambda$=$a$ (Fig.~\ref{fig4}(a)). But on increasing $x_{ns}$ further (Fig.~\ref{fig5}(b)), we immediately observe that $\langle{S_z^{tot}}\rangle$ decreases much more rapidly 
at lower temperatures followed by a slow decay toward the high temperatures. Also, in Fig.~\ref{fig5}(c) and (d), an inflection appears in $\langle{S_z^{tot}}\rangle$ around 
$T/T^*$$\sim$0.6. In this case $\langle{S_z^{in}}\rangle$ too exhibits a non-trivial behavior for all values of $x_{ns}$, which is unlike the 
case of $\lambda$=$a$. For example, for $x_{ns}$=0.05 (Fig.~\ref{fig5}(c)), there is a shoulder-like feature in $\langle{S_z^{in}}\rangle$ around 
$T/T^*$$\sim$0.05, which is absent for $\lambda$=$a$ (Fig.~\ref{fig4}(c)). Thus unlike the case of $\lambda$=$a$, where the intra-cluster couplings 
dominate, there are other relevant couplings, like the inter-cluster ones and those between the cluster impurities and bulk impurities, 
which come into play. On the other hand, the $\langle{S_z^{out}}\rangle$ curves are typically concave for all considered $x_{ns}$, and exhibit a long tail toward the higher 
temperatures.  They exhibit a sharp fall-off at low temperatures with increasing $x_{ns}$. At $T/T^*$$\sim$0.2, for $x_{ns}$=0.01, the value of $\langle{S_z^{out}}\rangle$ 
is about 0.8 which falls rapidly to almost 0.3 for $x_{ns}$=0.05. With increasing $x_{ns}$, the concentration of impurities outside the clusters gradually decreases, leading to 
an increase of the typical distance between them. Consequently the impurities outside interact more weakly with each other and this explains the sharp fall-off in 
$\langle{S_z^{out}}\rangle$ at lower temperatures.

\begin{figure}[htbp]
\centering
\subfigure{
\includegraphics[width=3.2in,angle=0]{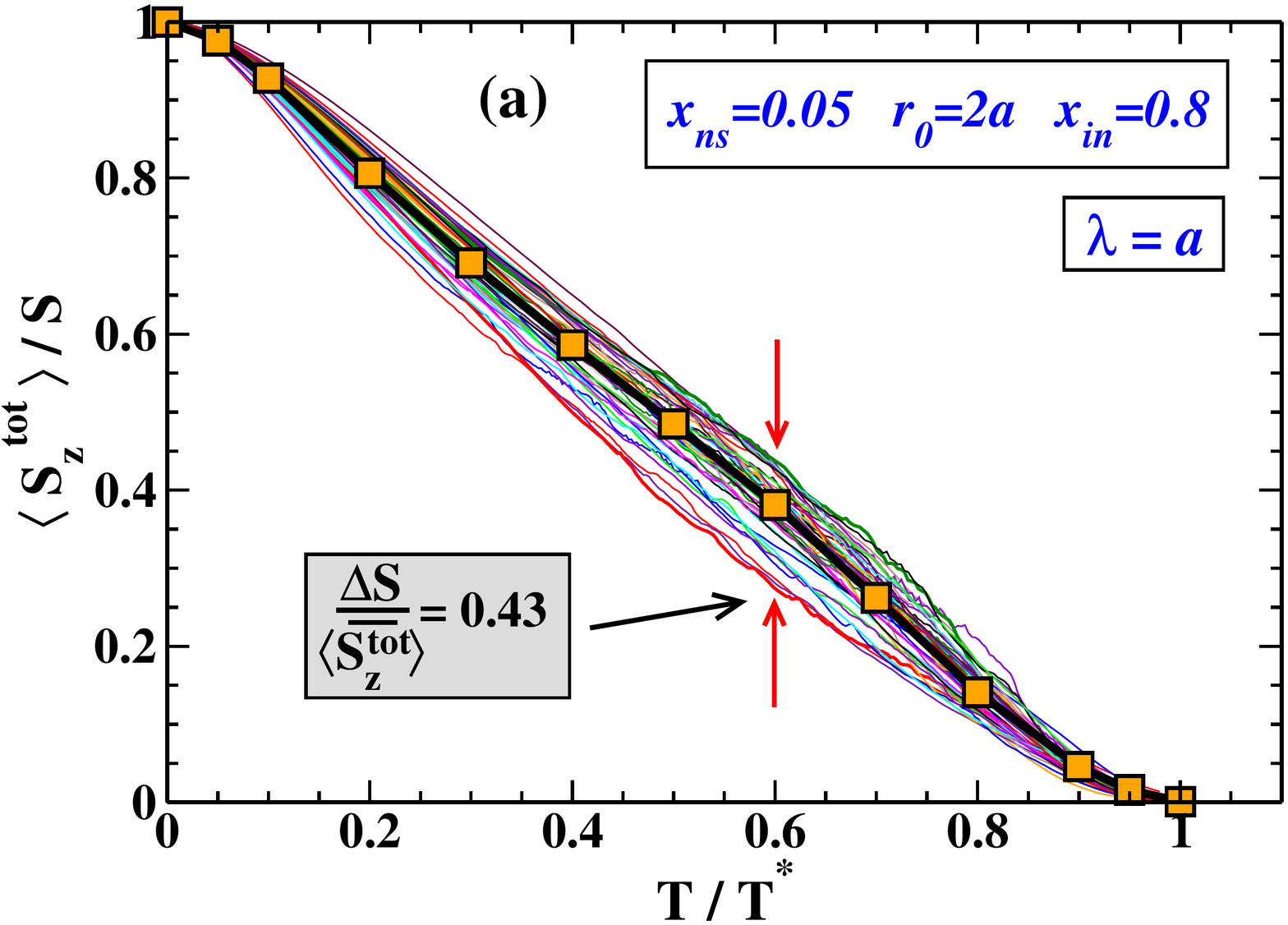}
\label{fig6a}
}
\hspace{-1.0cm}
\subfigure{
\includegraphics[width=3.2in,angle=0]{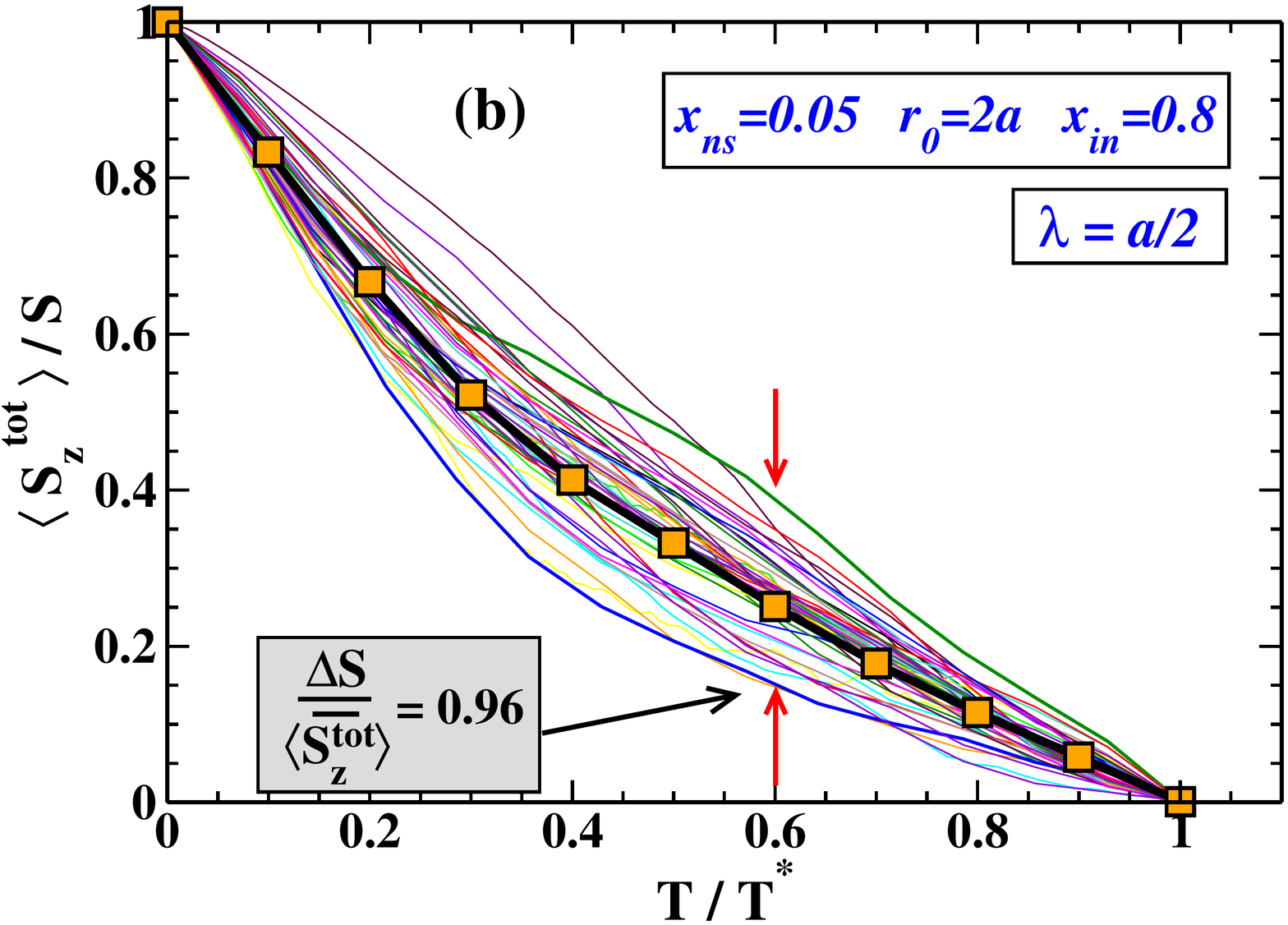}
\label{fig6b}
}
\label{fig6}
\caption[Optional caption for list of figures]{
%Caption of subfigures \ref{fig2a}, \ref{fig2b}
(Color online) Total average magnetization as a function of $T$/$T^*$ for $x_{ns}$=0.05 calculated for 50 configurations (thin continuous lines), for (a)$\lambda$=$a$, and 
(b)$\lambda$=$a/2$. The thick black lines with symbols indicate the configuration averaged magnetization ${\langle\overline{S_z^{tot}}\rangle}$. 
$\frac{\Delta S}{\langle \overline{S_{z}^{tot}}\rangle}$ is a measure of the fluctuation at $T/T^*$=0.6. (Here $r_0$=$2a$, $x_{in}$=0.8 and $N$=24$^3$).
}
\end{figure}

\ In the previous two figures we have discussed the cases for a single configuration of disorder only. Let us now analyze how the results depend on the random 
cluster configurations. Fig. 6 shows the average magnetization calculated for 50 configurations of disorder, for $x_{ns}$=0.05. We compare the case of the intermediate 
couplings to the short-ranged ones. One sees for the case of $\lambda$=$a$ (Fig.~\ref{fig6a}), 
that the magnetization curves have an almost linear similar shape over the entire temperature range. In fact the configuration averaged magnetization 
can be well approximated by $\langle{S_z^{tot}}\rangle$$\approx$$(1-T/T^*)$. On the other hand, the configuration averaged magnetization has a pronounced concave 
nature for $\lambda$=$a/2$ (Fig.~\ref{fig6b}). The decay slope at low temperatures ($T/T^*$$\le$0.2) is twice that of $\lambda$=$a$. Concerning the disorder fluctuations, one 
clearly sees that they are much stronger in the case of the short-ranged couplings. For example, the fluctuation of $\langle{S_z^{tot}}\rangle$ 
at $T/T^*$=0.6 is found to be more than doubled compared to that of $\lambda$=$a$.  For $\lambda$=$a/2$, most of the magnetization curves are concave in nature, while some 
exhibit a linear-like behavior. The typical separations between the clusters play a decisive role, in the presence of short-ranged couplings. This has been discussed 
in the context of Curie temperatures in inhomogeneous systems\cite{akash}.  Here it should be noted that the concave-like curves correspond to higher critical 
temperatures, while the linear ones coincide with relatively low $T_C$'s. Finally, we have found that for relatively 
 extended couplings ($\lambda$$\approx$2a) (not shown here), $\langle{S_z^{tot}}\rangle$ exhibits a more convex-like behavior. The fluctuations 
resulting from the disorder configurations were found to be much smaller than that of $\lambda$=$a$.

\begin{figure}[t]\centerline
{\includegraphics[width=5.5in,angle=0]{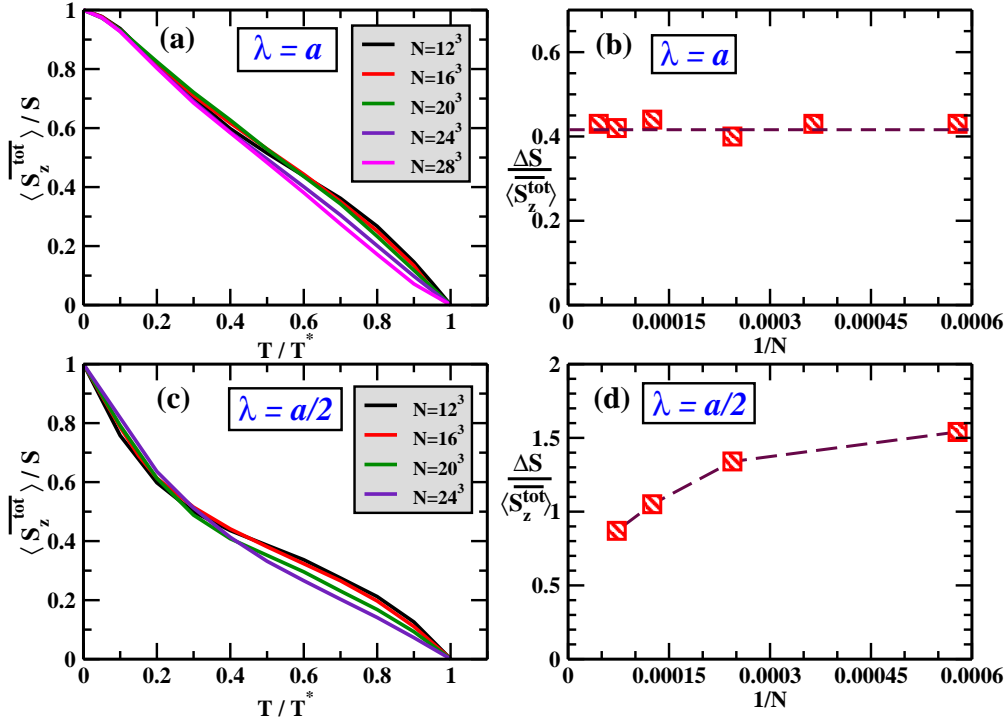}}
\caption{(Color online) Configuration averaged magnetization ${\langle\overline{S_z^{tot}}\rangle}$ for different system sizes $N$=$L^3$, 
corresponding to (a) $\lambda$=$a$, and (c) $\lambda$=$a/2$. The total number of configurations is of the order of few hundred. The respective
fluctuations $\frac{\Delta S}{\langle \overline{S_{z}^{tot}}\rangle}$ at $T/T^*$=0.6, as a function of $1/N$ is shown in (b) for $\lambda$=$a$, and 
(d) for $\lambda$=$a/2$. (Here $x_{ns}$=0.05, $r_0$=$2a$, and $x_{in}$=0.8). 
} 
\label{fig7}
\end{figure}

\ We now proceed further with the study of the finite size effects. Indeed, in inhomogeneous systems one naturally expects 
finite size effects to be much stronger than in homogeneously diluted ones.  These effects can be even more drastic for larger size of inhomogeneities.
The configuration averaged total magnetizations as a function of $T$/$T^*$, corresponding to $\lambda$=$a$ and $\lambda$=$a/2$, are shown respectively in 
Fig.~\ref{fig7}(a) and (c). The calculations are performed over different system sizes varying from $N$=12$^3$ up to $N$=28$^3$. $\langle\overline{S_z^{tot}}\rangle$ is 
the averaged magnetization obtained over a sample of few hundred disorder configurations. In both cases  we observe that 
the magnetization curves are very similar and the finite size effects are very weak. Let us now discuss the size dependence of the fluctuations, 
$\frac{\Delta S}{\langle \overline{S_{z}^{tot}}\rangle}$, (with respect to disorder configurations) of the average magnetization $\langle S_{z}^{tot}\rangle$.
Fig.~\ref{fig7}(b) and (d) show 
$\frac{\Delta S}{\langle \overline{S_{z}^{tot}}\rangle}$ at $T/T^*$=0.6 as a function of $1/N$, corresponding to $\lambda$=$a$ and $\lambda$=$a/2$, respectively. 
For $\lambda$=$a$, it is almost constant by varying the system sizes, with a value of around 0.42.
Interestingly for the short-ranged couplings, $\frac{\Delta S}{\langle \overline{S_{z}^{tot}}\rangle}$ decays monotonously, and 
 indicates that it saturates to a value of about 0.7 in the thermodynamic limit. It is important to note that for homogeneously diluted systems 
 $\frac{\Delta S}{\langle \overline{S_{z}^{tot}}\rangle}$  
vanishes in the thermodynamic limit. This means that the total average magnetization is not a self averaging quantity in the presence of inhomogeneities.

\begin{figure}[t]\centerline
{\includegraphics[width=6.0in,angle=0]{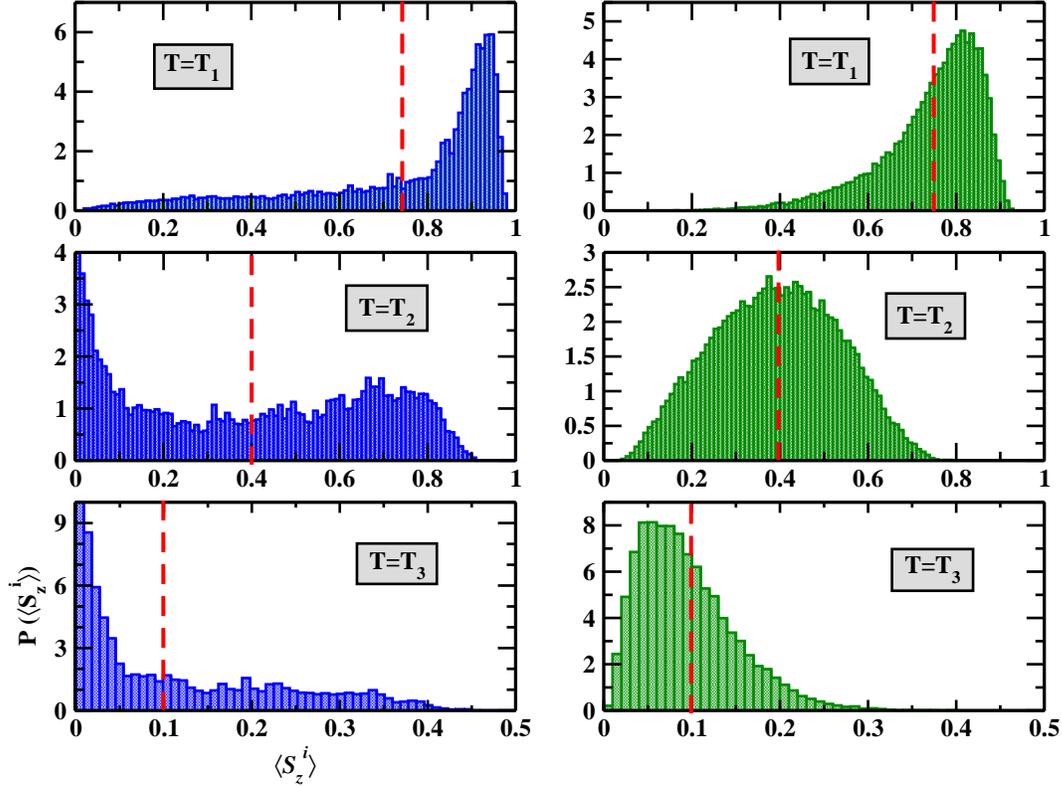}}
\caption{(Color online) Distribution of local magnetizations $\langle{S_z^{i}}\rangle$ for $\lambda$=$a/2$, at three different temperatures $T_1$, $T_2$, and $T_3$, 
which correspond to $\langle{S_z^{tot}}\rangle$=0.75$S$, 0.4$S$ and 0.1$S$ respectively.
Left column: Inhomogeneous case for $x_{ns}$=0.05 (with $r_0$=$2a$ and $x_{in}$=0.8). Right column: Homogeneous case. (The red dashed lines indicate the values of 
$\langle{S_z^{tot}}\rangle$/$S$ corresponding to the three different temperatures).
} 
\label{fig8}
\end{figure}

\ In order to get a deeper insight into the temperature dependence of the 
magnetization in the presence of inhomogeneities, we analyze the nature of the distributions of the local magnetizations. 
In Fig.~\ref{fig8}, we compare the distributions of $\langle{S_z^{i}}\rangle$'s in the inhomogeneous systems with those of the 
homogeneously diluted ones, for $\lambda$=$a/2$.  The distributions are shown for three different temperatures $T_1$, $T_2$, 
and $T_3$, which correspond to  $\langle{S_z^{tot}}\rangle$=0.75$S$, 0.4$S$, and 0.1$S$, respectively. The distributions at 
the lowest temperature $T_1$ are almost similar for both cases, except for an extended tail toward the small magnetization values 
in the inhomogeneous system.  At the intermediate temperature $T_2$, the distributions reveal a more pronounced difference. It has a Gaussian-like shape 
for the homogeneous systems, whereas it has a clear bimodal nature in the inhomogeneous ones. It is also wider than that of the homogeneous case. 
There is a significantly higher weight below $\langle{S_z^{i}}\rangle$=0.2 for the inhomogeneous systems. This relatively high weight corresponds to the impurities outside and far from the 
 clusters. The broad peak around $\langle{S_z^{i}}\rangle$$\approx$0.7 is associated with the impurities inside the clusters. Finally, for the highest temperature 
$T_3$, the distribution in the homogeneous systems is unimodal with a small tail at large $\langle{S_z^{i}}\rangle$'s and the half-width is of the order of 
$\langle{S_z^{tot}}\rangle$=0.1. On the other hand, in the inhomogeneous compounds, we observe a very strong weight for $\langle{S_z^{i}}\rangle$$\le$0.05 and an  
extended tail which goes up to 4$\langle{S_z^{tot}}\rangle$. For $T$=$T_3$, an estimate of the percentage of impurities within the range of 
0.05$\le$$\langle{S_z^{i}}\rangle$$\le$0.15 gives 65\% for the homogeneous systems and only 15\% in the presence of inhomogeneities. To conclude this discussion, 
we provide in Fig. 9 a real space illustration of the variation of the local magnetizations. The 2D-snapshots, are shown for both kinds of systems, at the 
temperatures $T_1$, $T_2$, and $T_3$, corresponding to $\lambda$=$a/2$.

\begin{figure}[htbnp]
\centering
\subfigure{
\includegraphics[width=5.2in,angle=0]{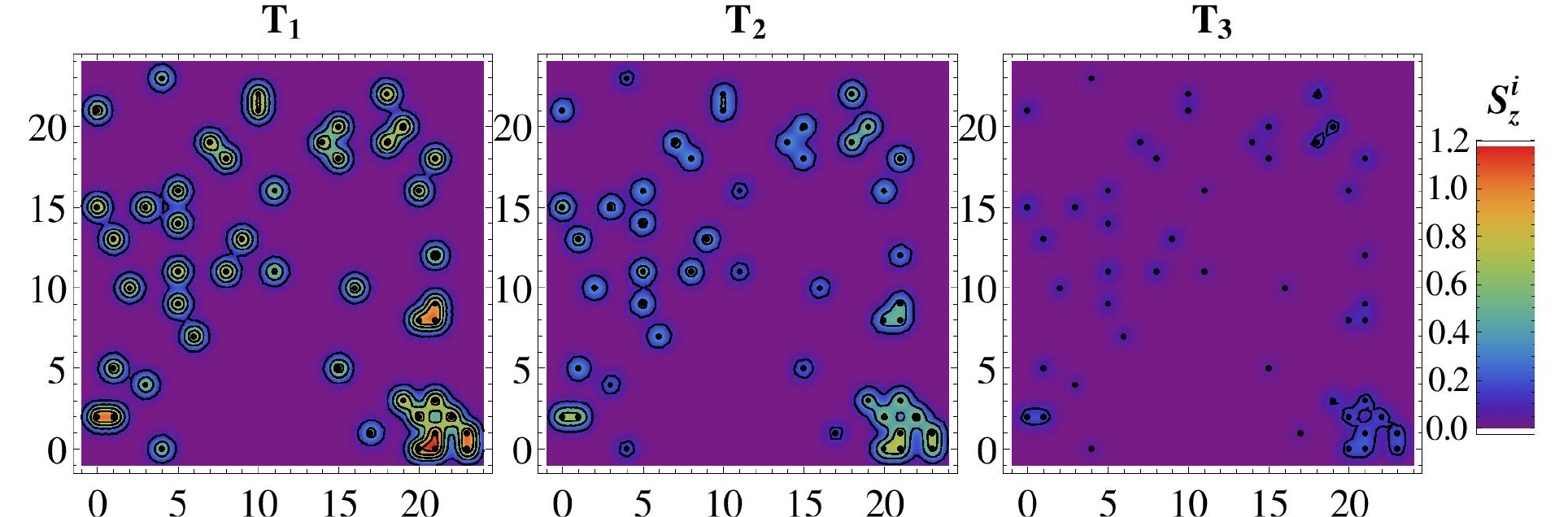}
\label{fig9a}
}
\subfigure{
\includegraphics[width=5.2in,angle=0]{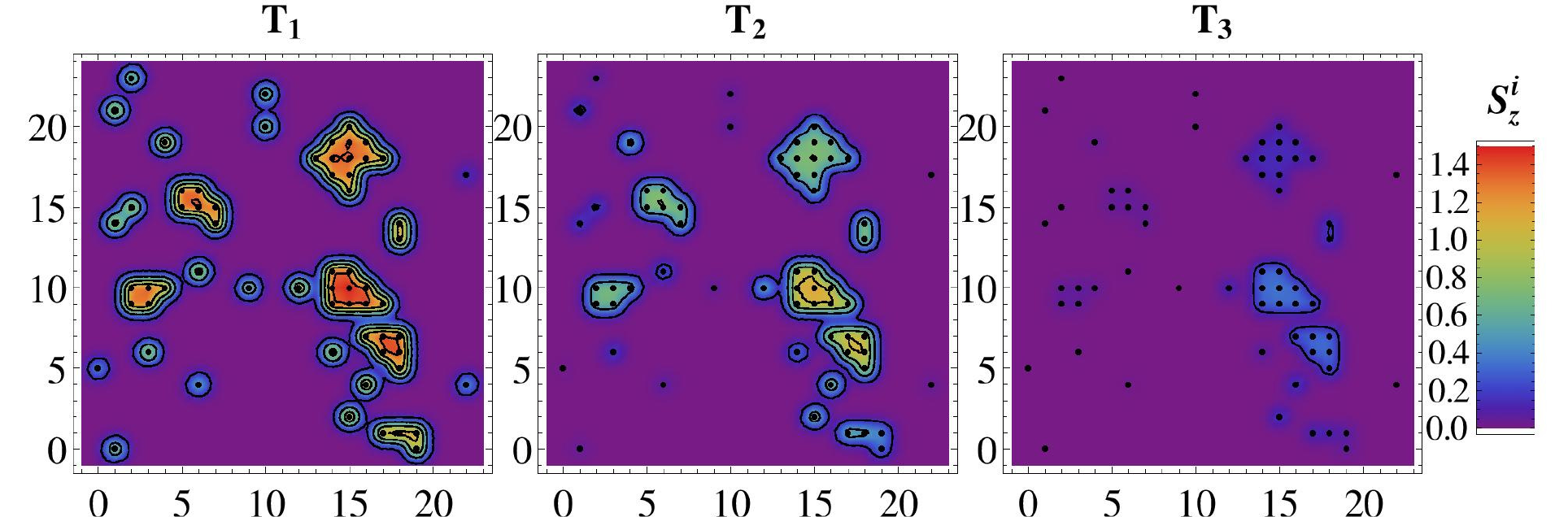}
\label{fig9b}
}
\label{fig9}
\caption[Optional caption for list of figures]{
%Caption of subfigures \ref{fig2a}, \ref{fig2b}
Snapshots of the local magnetizations in a 2D-plane for $\lambda$=$a/2$ . The top panel corresponds to the homogeneous case, and the bottom panel 
 to the inhomogeneous case for $x_{ns}$=0.05. $T_1$, $T_2$, and $T_3$ denote the temperatures when $\langle{S_z^{tot}}\rangle$=0.75$S$, 0.4$S$ and 
0.1$S$ respectively. (Here $N$=24$^3$). 
}
\end{figure}

\ Thus, to summarize, we have presented a detailed and comprehensive study of the effects of nanoscale inhomogeneities on the spontaneous magnetization in diluted 
magnetic systems, which had hitherto been unexplored. We have compared the average magnetization behavior in inhomogeneous systems to that of the homogeneously diluted case, 
for different ranges of the magnetic couplings. Unlike the convex nature of the spontaneous magnetization in homogeneous systems  
a linear temperature dependence (over the entire temperature range) is obtained in the inhomogeneous compounds for intermediate couplings. Whereas it exhibits a pronounced 
concave shape in systems with 
short-ranged interactions. Additionally, the local magnetizations show bimodal and wider distributions in contrast to that of the homogeneously diluted systems, where it 
remains unimodal. The finite 
size analyses have revealed that the fluctuations of the average magnetization (with respect to disorder configurations) remain finite in the thermodynamic limit 
 in inhomogeneous systems, unlike the homogeneously diluted case for which it vanishes. We believe it would be of great interest to corroborate our findings 
 by future experimental studies.

\acknowledgments
This work was supported by the EU within FP7-PEOPLE-ITN-2008, Grant number 234970 Nanoelectronics: Concepts, Theory and Modelling. We would like to thank S. Kettemann 
for useful discussions and insightful comments.

%\clearpage

\end{document}